\documentclass[referee]{raa}

\newcommand{\msunh}{\>h^{-1}\rm M_\odot}
\newcommand{\Msun}{\>{\rm M_\odot}}

\newcommand{\mpch}{\>h^{-1}{\rm {Mpc}}}

\newcommand{\kmsmpc}{\>{\rm km}\,{\rm s}^{-1}\,{\rm Mpc}^{-1}}
\newcommand{\Mstar}{M_{\ast}}
\newcommand{\Mh}{M_{\rm h}}
\newcommand{\Mhi}{M_{\rm HI}}
\newcommand{\umag}{\>^{0.1}{\rm M}_u-5\log h}

\usepackage{graphicx,times}
\usepackage{xcolor}
\usepackage{natbib}
\usepackage{amssymb,amsmath}
\bibpunct{(}{)}{;}{a}{}{,}

\usepackage[a4paper=true,dvipdfm=true,pagebackref=true]{hyperref}
\hypersetup{pdftitle = The title of my PDF, pdfauthor = My name, pdfsubject= The subject, pdfkeywords = keyword1 keyword2 keyword3}
\hypersetup{colorlinks = true, linkcolor = green, anchorcolor = red, citecolor = blue, filecolor = red, pagecolor = red, urlcolor = red}

\begin{document}

   \title{An empirical model to form and evolve galaxies in dark matter halos}

 \volnopage{ {\bf 2012} Vol.\ {\bf X} No. {\bf XX}, 000--000}
   \setcounter{page}{1}

\author{Shijie Li\inst{1,6}, Youcai Zhang\inst{1},
  Xiaohu Yang\inst{2,3}, Huiyuan Wang\inst{4}, Dylan
  Tweed\inst{2}, Chengze Liu\inst{2}, Lei
  Yang\inst{2}, Feng Shi\inst{1,6}, Yi Lu\inst{1}, Wentao
  Luo\inst{1}, Jianwen Wei\inst{5}}

   \institute{  Key Laboratory for Research in Galaxies and
  Cosmology, Shanghai Astronomical Observatory; Nandan Road 80,
  Shanghai 200030, China; {\it sjli@shao.ac.cn} \\
    \and
        Center for Astronomy and Astrophysics, Shanghai Jiao
  Tong University, 800 Dongchuan Road, Shanghai 200240, China; {\it xyang@sjtu.edu.cn} \\
    \and
        IFSA Collaborative Innovation Center, Shanghai Jiao
  Tong University, 800 Dongchuan Road, Shanghai 200240, China\\
    \and
        Key Laboratory for Research in Galaxies and
  Cosmology, Department of Astronomy, University of Science and
  Technology of China, Hefei, Anhui 230026, China\\
    \and
        Center for High Performance Computing, Shanghai Jiao
  Tong University, 800 Dongchuan Road, Shanghai 200240, China\\
    \and
        University of Chinese Academy of Sciences, 19A, Yuquan Road,
    Beijing, China\\
   \vs \no
   {\small Received XXXX ; accepted XXXX}
  }

  \abstract{
  Based on the star formation histories (SFH) of galaxies in halos of  different
  masses, we develop  an  empirical model to grow galaxies in dark mattet halos.
  This  model  has very  few  ingredients, any  of  which  can  be associated to
  observational data and thus be efficiently assessed. By applying this model to
  a very high resolution cosmological $N$-body simulation, we predict  a  number
  of galaxy properties that are a very good match to relevant observational data.
  Namely, for  both  centrals and satellites, the galaxy stellar  mass  function
  (SMF)  up  to redshift $z\simeq4$ and the conditional stellar  mass  functions
  (CSMF)  in  the  local  universe  are in good  agreement with observations. In
  addition, the 2-point  correlation  is well predicted in the different stellar
  mass  ranges  explored  by  our  model. Furthermore,  after   applying stellar
  population   synthesis   models   to  our stellar composition as a function of
  redshift, we  find  that  the  luminosity  functions  in $^{0.1}u$, $^{0.1}g$,
  $^{0.1}r$, $^{0.1}i$ and $^{0.1}z$  bands  agree   quite  well  with  the SDSS
  observational results down to an  absolute magnitude at about -17.0. The  SDSS
  conditional  luminosity  functions (CLF) itself  is  predicted  well. Finally,
  the  cold  gas is derived from the star formation rate (SFR) to predict the HI
  gas mass within each mock galaxy. We find  a remarkably good match to observed
  HI-to-stellar mass ratios. These features ensure that such galaxy/gas catalogs
  can be used to generate reliable mock redshift surveys.
}

   \authorrunning{S.-J. Li et al. }
   \titlerunning{Empirical model of galaxy formation}
   \maketitle

%________________________________________________ sections below
%
\section{Introduction}
\label{sect:intro}

Galaxies are thought to form and evolve in cold dark matter (CDM) halos, however,
our understanding of the galaxy formation mechanisms and the interaction between
baryons and dark matter are still quite poor, especially quantitatively
(see \citealt{2010gfe..book.....M}, for a detailed review). Within  hydrodynamic
cosmological simulations, the evolution of the gas component is described on top
of the dark matter, with extensive implementation of cooling, star formation and
feedback processes. Such  detailed  implementation  of galaxy formation within a
cosmological framework requires vast computational time and resources
(\citealt{2005Natur.435..629S}).

However the formation of dark matter halos can be easily derived and interpreted,
such merger trees can be derived directly from $N$-body simulations, or  through
Monte Carlo methods. Within those trees, sub-grid  models  can be applied on the
scale of DM halos themselves. Such  models  are referred as semi-analytic models
(hereafter SAM), and provide the means to test galaxy formation models at a much
lower computational cost  (\citealt{2007MNRAS.377...63C}).

In SAMs, some simple equations describing  the  underlying  physical ingredients
regarding the accretion and cooling of gas, star formation etc..., are connected
to the dark  matter  halo  properties, so that the baryons can evolve within the
dark matter halos  merger  trees. The related free parameters in these equations
are tuned to statistically match some physical properties of observed galaxies.

The basic principles of modern SAMs were first introduced by
\cite{1991ApJ...379...52W}. Consequently  numerous  authors participated  in the
studies of such models and made great progresses (e.g. \citealt{1993MNRAS.264..201K};
\citealt{1998MNRAS.295..319M}; \citealt{1999MNRAS.310.1087S};
\citealt{2000MNRAS.319..168C}; \citealt{2004MNRAS.348..333D};
\citealt{2005ApJ...631...21K}; \citealt{2006MNRAS.365...11C};
\citealt{2006MNRAS.370..645B}; \citealt{2007MNRAS.375.1189M};
\citealt{2011MNRAS.413..101G}). Through     the  steerable  parameters, SAM  has
reproduced  many  statistical  properties  of large galaxy samples in  the local
universe  such  as  luminosity  functions, galactic   stellar   mass  functions,
correlation functions, Tull-Fisher relations, metallicity-stellar mass relations,
black hole-bulge mass relations and color-magnitude relations. However, the main
shortcoming of SAMs is that there are too many free parameters and degeneracies.
Despite  the  successes  of these galaxy formation models, the sub-grid  physics
is still poorly  understood (\citealt{2012NewA...17..175B}). By  tuning the free
parameters, the SAM prediction could match some of the observed galaxy properties
in consideration, especially in the local universe. But none of the current SAMs
can match the low and high redshift data simultaneously (\citealt{2012MNRAS.423.1992S}).
Traditionally, parameters are preferably set without providing a clear statistical
measure of success for a combination of observed galaxy properties.

As  a  SAM  cost  much  less  computation time than a full hydrodynamical galaxy
formation simulation, one is allowed to explore a  wide range of parameter space
in acceptable time interval. To better constrain the SAM parameters, Monte Carlo
Markov Chains (MCMC) method  has been applied to SAMs in recent years. The first
paper that  incorporated MCMC into SAM is \cite{2008MNRAS.384.1414K}, which used
the  star  formation  rate  and  metallicity as model constraint. Some other SAM
groups also have developed their own models associated with the MCMC method (e.g.
\citealt{2009MNRAS.396..535H,2013MNRAS.431.3373H};\citealt{2010MNRAS.405.1573B};
\citealt{2010MNRAS.407.2017B}; \citealt{2011MNRAS.416.1949L,2012MNRAS.421.1779L};
\citealt{2013MNRAS.428.2001M}). The details of MCMC are beyond the  aims of this
paper, we refer the readers to these relevant literatures
(\citealt{2007nrca.book.....P}; \citealt{2008ConPh..49...71T}).

As  pointed  out  in  \cite{2010MNRAS.405.1573B},  our  understanding  of galaxy
formation  is  far  from  complete. SAMs should not be thought of as attempts to
provide  a  final  theory  of galaxy formation, but instead to provide a mean by
which  new  ideas  and  insights  may  be  tested  and by which quantitative and
observationally comparable predictions may be extracted in order to test current
theories. Because  of the  large number of free parameters, new ideas and sights
relevant  with  the  sub-grid   physics  may  often  bring new degeneracies with
increased  complexity  and  uncertainties  to  the  model either traditional SAM
or MCMC. In  general, if we take a step back from SAMs, we find that the largest
part  of  the  parameters  and uncertainties are related to the sub-grid physics
implemented  for  the gas. Focussing the model on the formation and evolution of
the {\it stars} within dark matter halos, the vast majority of the uncertainties
in SAM related with the gas component will be reduced.

Understanding  the  relation  between  dark matter halos and galaxies is a vital
step to model galaxy formation and evolution in  dark  matter  halos.  In recent
years, we  have  seen  drastic  progress  in establishing the connection between
galaxies and dark matter halos, such  as  the halo occupation distribution (HOD)
models (e.g. \citealt{1998ApJ...494....1J}, \citealt{2002ApJ...575..587B},
\citealt{2005ApJ...630....1Z}, \citealt{2010MNRAS.406..147F},
\citealt{2011ApJ...738...22W}, \citealt{2012ApJ...751L..44W},
\citealt{2012ApJ...744..159L}), and the closely related conditional stellar mass
(or luminosity) function models(\citealt{2003MNRAS.339.1057Y},
\citealt{2003MNRAS.340..771V}, \citealt{2006ApJ...647..201C},
\citealt{2007MNRAS.376..841V}, \citealt{2009ApJ...695..900Y},
\citealt{2012ApJ...752...41Y}, \citealt{2015ApJ...799..130R}).  The  former make
use of the clustering of galaxies  to  constrain  the probability of finding $N$
galaxies in a halo of mass $M$. While the latter make use of both clustering and
luminosity(stellar mass)  functions  to  constrain  the   probability of finding
galaxies with given  luminosity (or stellar mass) in  a  halo  of mass $M$. In a
recent study, \cite{2012ApJ...752...41Y}(hereafter Y12) proposed a self-consistent
model properly taking into account (1) the  evolution  of  stellar-to-halo  mass
relation   of  central  galaxies; (2) the  accretion and subsequent evolution of
satellite galaxies. Based on the host halo and subhalo accretion models provided
in \cite{2009ApJ...707..354Z} and \cite{2011ApJ...741...13Y}, Y12  obtained  the
conditional  stellar  mass  functions (CSMFs)  for  both  central  and satellite
galaxies  as  functions  of  redshift. Based  on  the mass assembly histories of
central galaxies, the amount of accreted  satellite galaxies and the fraction of
surviving satellite galaxies  constrained in Y12, we obtained the star formation
histories (SFH) of central galaxies in halos of different masses
(\citealt{2013ApJ...770..115Y}). Similar  SFH models were also proposed based on
$N$-body or Monte Carlo merger trees (e.g., \citealt{2013MNRAS.428.3121M};
\citealt{2013ApJ...770...57B}; \citealt{2014MNRAS.439.1294L}). These   SFH  maps
give us the   opportunity   to grow galaxies in $N$-body simulations without the
need to model the complicated gas physics. In those models referred as empirical
model (EM) of   galaxy   formation, the   growth   of  galaxies is statistically
constrained using observational data.

This   paper  is  organized  as follows. In section 2, we describe in detail our
simulation  data  and  EM  model. In  section 3, we  show  our model predictions
associated  with  the  stellar masses of galaxies. The model predictions related
with the luminosity and HI gas components are presented in sections 4.  Finally,
in section 5, we make our  conclusions and discuss the applications of our model
and the galaxy catalog thus constructed.

\section{Simulation and our empirical model}
\label{sec_model}

\subsection{The simulation}

\begin{figure}
\centering
\includegraphics[width=8.5cm]{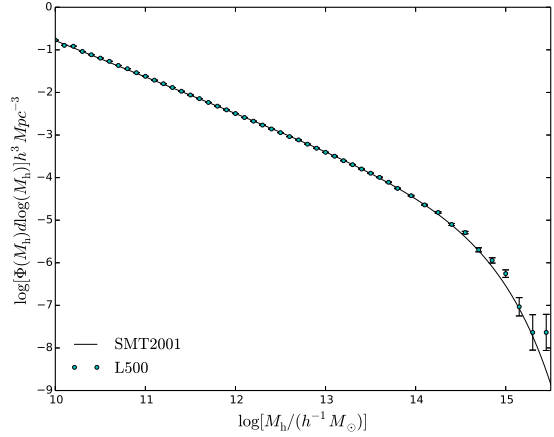}
\caption{Halo mass functions of the simulation. The black curve and cyan circles
  represent   respectively   the \cite{2001MNRAS.323....1S} (SMT2001)   analytic
  prediction and data extracted from the L500 simulation. }
\label{fig:hmf}
\end{figure}

Similar to the SAMs, our EM also starts from dark matter  halo  merger trees. In
this study we use dark matter halo merger trees extracted from a high resolution
$N$-body simulation. The  simulation  describes the evolution of the phase-space
distribution of $3072^{3}$ dark matter particles in a periodic box of $500 \mpch$
on a side. It  was carried out in  the Center  for  High  Performance Computing,
Shanghai Jiao Tong University. This simulation, hereby referred as L500, was run
with {\tt L-GADGET}, a memory-optimized version of {\tt GADGET2}
(\citealt{2005Natur.435..629S}). The   cosmological  parameters  adopted by this
simulation are consistent with WMAP9 results as follows: $\Omega_{\rm m} = 0.282$,
$\Omega_{\Lambda} = 0.718$, $\Omega_{\rm b} = 0.046$, $n_{\rm s}=0.965$,
$h=H_0/(100 \kmsmpc) = 0.697$ and $\sigma_8 = 0.817$ (\citealt{2013ApJS..208...19H}).
The particle masses and softening lengths are, respectively, $3.3747\times10^{8}\msunh$
and $3.5 h^{-1}\rm kpc$. The  simulation  is started at redshift 100 and has 100
outputs from z=19, equally spaced in $\log$ (1 + z).

Dark matter halos were first identified by the friends-of-friends(FOF) algorithm
with linking length of $0.2$ times the mean  particle  separation and containing
at least $20$ particles.  The corresponding dark matter  halo mass function (MF)
of  this  simulation  at  redshift  $z=0$  is  represented  by  cyan  circles in
Fig.~\ref{fig:hmf}, while the  black  curve  corresponds  to  the analytic model
prediction  by  \cite{2001MNRAS.323....1S}(SMT2001). The  halo  mass function of
this  simulation  is in good agreement with the analytic model prediction in the
related mass ranges.

Based on halos at different outputs, halo merger trees were constructed
\cite{1993MNRAS.262..627L}. We first use the SUNFIND algorithm
(\citealt{2001MNRAS.328..726S}) to identify the  bound substructures  within the
FOF halos or FOF groups. In a FOF group, the most massive substructure is defined
as main halo and the other substructures are defined as subhalos. Each  particle
contained in a given subhalo or main halo is assigned a  weight  which decreases
with  the  binding  energy.  We  then  find  all  main halos and subhalos in the
subsequent snapshot that contain some  of  its  particles. The descendant of any
(sub)halo  is  chosen  as  the  one  with  the  highest weighted count of common
particles. This  criteria  can  be  understood as a weighed maximum shared merit
function (see \citealt{2005Natur.435..629S} for  more  details).  Note that, for
some small halos, the tracks of which are temporarily lost in subsequent snapshot,
we skip one snapshot in finding their descendants. These descendants are called
``non-direct descendant".

\subsection{The empirical model of galaxy formation}
\label{sec_sfr}

Unlike any SAM where each halo initially gets a lump of hot gas to be eventually
turned into a galaxy (\citealt{2006RPPh...69.3101B}), our EM starts  with stars.
Here we make use of the SFH map of dark matter halos obtained by
\cite{2013ApJ...770..115Y} to grow galaxies. In our EM of galaxy formation,
\emph{central} and \emph{satellite} galaxies   are  assumed to be located at the
center of the main halos and subhalos respectively. Their velocities are assigned
using  those  of the main halos and subhalos. For those satellite galaxies whose
subhalos  are disrupted, (e. g. orphan galaxies) the  host  halo  is   populated
according to its NFW profile. Their   velocities  are  assigned according to the
halo velocity combined with the velocity dispersion (see \citealt{2004MNRAS.350.1153Y}
for the details of such an assignment).

Apart from the obvious issue of positioning mock galaxies, we have to  implement
the  stellar  mass  evolution. For  central and satellite galaxies, stellar mass
$M_{\star, c}(t_2)$ at a time $t_2$ is  derived  by  adding  to the stellar mass
$M_{\star, c}(t_1)$ at a time $t_1$ the contribution from star formation
$\Delta M_{\star,c}(t_1)$ and disrupted satellites  $\Delta M_{\star, dis}(t_1)$
as follows:

\begin{equation}\label{eq:central}
M_{\star}(t_2) = M_{\star}(t_1) + \Delta M_{\star}(t_1,t_2) +
  \Delta M_{\star, dis}(t_1,t_2)
\end{equation}

Obviously before implementing these models, the galaxies have to  be seeded. For
each halo and subhalo, we follow the merger tree back in  time  to determine the
earliest  time  output (at $t_{\rm min}$) when  it  was identified as a halo (at
least 20 particles). Then a seed galaxy with initial stellar mass
$ M_{\star}(t_{\rm min})$ is  assigned  to  this halo at the beginning redshift.
Here  the  stellar  mass  is  assigned  according  to the central-host halo mass
relation obtained by \cite{2012ApJ...752...41Y}, taking into account the cosmology
of our simulations. We note that only halos with direct descendants are seeded.

\subsubsection{Star formation of central galaxies}

We first model the growth of {\it central} galaxies that are associated with the
host (main) halos. Listed below are the details.
\begin{itemize}

\item In order to integrate the contribution of star formation between snapshots
  corresponding  to  times  $t_1$ and $t_2=t_1+\Delta T$, we  increase  the time
  resolution by defining smaller timesteps $\Delta t=\Delta T/N$. Here we choose
  $N=5$, since greater values have very limited impact  on  the results. We also
  assume that the SFR is constant during any time step $\Delta t$.

\item Then  we  estimate ${\dot M}_{\star}(t)$ the SFR of central galaxy at time
  $t$ in  a halo  with mass $M_{\rm h}$. As shown in \cite{2013ApJ...770..115Y},
  the distribution  of  SFR of central galaxies have quite large scatters around
  the  median  values and show quite prominent bimodal features.  To partly take
  into account  these scatters, for each timestep $\Delta t$, the star formation
  rate  ${\dot M}_{\star}(t)$ is   drawn  from  a lognormal distribution of mean
  ${\dot M}_{\star, 0}(t)$ and    dispersion $\sigma$. So   the  SFR  of central
  galaxies is indeed set as:
\begin{equation} \label{eq:SFR}
  \log   {\dot M}_{\star}(t) = \log {\dot M}_{\star, 0}(t) +
      \sigma \cdot N_{\rm gasdev} \, ,
  \end{equation}
  where ${\dot M}_{\star, 0}(t)$ is the median SFR predicted by
  \cite{2013ApJ...770..115Y} and  $N_{\rm gasdev}$ is  a random number generated
  using  the  code  of Numerical Recipe(\citealt{2007nrca.book.....P}). Here  we
  adopt a $\sigma=0.3$ lognormal scatter as suggested in \cite{2013ApJ...770..115Y}.

\item The stellar mass formed between the snapshots, $\Delta
  M_{\star}(t_1,t_2)$ is determined as:
  \begin{equation} \label{eq:deltam}
   \Delta M_{\star}(t_1,t_2) =  \sum_{t=t_1}^{t=t_2} {\dot M}_{\star}(t)
      \Delta t \, .
  \end{equation}

\end{itemize}

\subsubsection{Star formation of satellite galaxies}
After  focussing  on  the  growth  of  central galaxies, we need to focus on the
satellite  galaxies. We  start  by  modeling  their  growth while they are still
associated  to  subhalos. Once  the  host halo enters a bigger one and becomes a
subhalo, the  SFR  of  the new satellite is expected to decline as a function of
time due to the stripping effect, etc. Here we use  the  star formation model of
satellite galaxy proposed by \cite{2014MNRAS.439.1294L} to  construct their star
formation history. A  simple $\tau$ model  has been adopted to describe the star
formation rate decline in \cite{2014MNRAS.439.1294L} as follows:
\begin{equation} \label{eq:sfrsat}
{\dot M}_{\star,{\rm  sat}}(t) = {\dot M}_{\star}(t_{a})
		\exp\left( -\frac{t-t_{a}}{\tau_{\rm sat}} \right) \, ,
\end{equation}
where $t_{a}$ is the time  when the galaxy is accreted into its host to become a
satellite and ${\dot M}_{\star}(t_{a})$ the corresponding SFR.  $\tau_{\rm sat}$
is  the  exponential  decay  time  scale  characterizing the decline of the star
formation for a galaxy of stellar mass $M_{\star}$. We adopt the following model
of the characteristic time
\begin{equation} \label{eq:sfrsat_t}
\tau_{\rm sat} = \tau_{\rm sat,0} \exp\left( -\frac{M_\star}{M_{\star,c}} \right) \, ,
\end{equation}
where  $\tau_{\rm sat,0}$  is  the  time  for  a  galaxy  with a stellar mass of
$M_{\star,c}$. The values $\tau_{\rm sat,0}$ and $M_{\star,c}$ used in our model
are the best fit values of MODEL III in \cite{2014MNRAS.439.1294L} with
$\log (H_0 \tau_{\rm sat,0})=-1.37$ and $\log M_{\star,c}=-1.4$.

The growth of the satellite stellar mass is thus becomes:

\begin{equation} \label{eq:deltam_sat}
   \Delta M_{\star,sat}(t_1,t_2) = {\dot M}_{\star}(t_{a}) \exp\left(
     -\frac{t_2-t_{a}}{\tau_{\rm sat}} \right) \cdot \Delta T
\end{equation}

\subsubsection{Merging and stripping of satellite galaxies}
\label{sec_merger}

Apart  from  the {\it in situ} star  formation, another important process in our
model  is  the  merging and stripping of satellite galaxies. The merging process
has been studied by many people through hydrodynamical simulation (e.g.
\citealt{2005ApJ...624..505Z}; \citealt{2008MNRAS.383...93B};
\citealt{2008ApJ...675.1095J}). Here   we  assume  that  the  satellite galaxies
orbiting  within  a  dark matter halo may experience dynamical friction and will
eventually be disrupted, while only a small fraction of stars are finally merged
with center galaxy of the halo.

So when a satellite cannot be associated with a subhalo, we use a delayed merger
scheme where the satellite coalless with the central after the dynamical friction
timescale described in the fitting formula of \cite{2008ApJ...675.1095J}:
\begin{equation}\label{eq:disrupt}
T_\mathrm{dyn}
= 1.4188\frac{r_{\rm c}M_{\rm h}}{v_{\rm c}M_{\rm sub}}
		\frac{1}{\ln(1+\frac{M_{\rm h}}{M_{\rm sub}})}	\,,
\end{equation}
where  $M_{\rm sub}$  and  $M_{\rm h}$ are  the  respective  {\it halo}   masses
associated to satellite and central galaxies, at the timestep a satellite galaxy
was last found in a subhalo. This formula is valid for a small satellite of halo
mass $M_{\rm sub}$ orbiting  at  a  radius $r_\mathrm{c}$ in  a halo of circular
velocity $v_\mathrm{c}$. As the satellite galaxy was last found in a  subhalo is
disrupted after $\Delta t=T_\mathrm{dyn}$, we transfer a fraction of its stellar
mass to the central galaxy. So that tha contribution of disrupted satellite follows
\begin{equation}\label{eq:merger}
\Delta M_{\star, dis}(t_1,t_2) = f_\mathrm{merger} \sum
  M_{\star, sat}(t_\mathrm{sat})\,  ,
\end{equation}
where $ M_{\star, sat}(t_{sat})$ is stellar  mass of the in-falling  satellite a
determined    when   it   was  last  found in a subhalo at $t_\mathrm{sat}$ with
$t_{sat}+T_\mathrm{dyn} \leq t_2$. $f_{merger}$ is  fraction  of  the  satellite
galaxy  stellar  mass merged into central galaxy.  Here $f_\mathrm{merger}=0.13$
is set to the best fit value of MODEL III in
\cite{2014MNRAS.439.1294L}.

\subsubsection{Passive evolution of galaxies}
\label{sec_merger}

Finally, we take into account the passive evolution of both central
and satellite galaxies. As we have the stellar mass composition of
each galaxy as a function of time, the final stellar mass is
determined as:
\begin{equation} \label{eq:finalm}
M_{\star}(t_0) =  M_{\star}(t_{\rm min}) \cdot f_{\rm passive}(t_0-t_{\rm
  min}) + \sum_{t=t_{\rm min}}^{t=t_0}\Delta {M}_{\star}(t)
      \cdot f_{\rm passive}(t_0-t) \, ,
\end{equation}
where $f_{\rm passive}(t)$ is the mass fraction of stars remaining at
time $t$ after the formation. We obtained $f_{\rm passive}(t)$ from
\cite{2003MNRAS.344.1000B}, courtesy of Stephane Charlot (private
communication).

\subsection{Other star formation history models}

There have been many other star formation history models proposed in recent years
(e.g. \citealt{2009ApJ...696..620C}, \citealt{2013ApJ...770...57B}). Here we make
use of the model constrained by \cite{2014MNRAS.439.1294L}, in  order to further
test our empirical model. This  model  is similar in a sense that it consists on
predicting SFR within halos and subhalos to build galaxies. In  the  larger part
of the result section, the properties of the central and satellite  galaxies are
compared to our fiducial EM predictions.

\cite{2014MNRAS.439.1294L} (hereafter Lu14)  developed  an empirical approach to
describe  the  star  formation  history  model of central galaxies and satellite
galaxies.  They assumed an analytic formula for the SFH of central galaxies with
a few free parameters. The galaxies grow in dark  matter halos based on the halo
merger trees generated by Extended Press-Schechter (EPS:
\citealt{1991ApJ...379..440B,1991MNRAS.248..332B}) formalism  and  Monte   Carlo
method. With different observation constraint, they got four different empirical
models. Here we  only pick Model III in Lu14 to compare with our model. In Lu14,
the star formation rate of central galaxies can be written as follows:
\begin{equation}\label{lu_cen}
  {\dot M}_\star =
    {\cal E} \frac{f_B M_{\rm vir}}{\tau_0} \left( 1+z \right)^{\kappa}
      (X+1)^{\alpha}
    \left(\frac{X+\mathcal{R}}{X+1}\right)^{\beta}
    \left(\frac{X}{X+\mathcal{R}}  \right)^{\gamma}
    \,,
\end{equation}
where ${\cal E}$ is an overall efficiency; $f_B$ is  the  cosmic  baryonic  mass
fraction; $\tau_0$ is a dynamic timescale of the halos  at  the present day, set
to be $\tau_0\equiv 1/(10 H_0)$; and $\kappa$ is  fixed  to  be ${3/2}$ so  that
$\tau_0/(1+z)^{3/2}$ is  roughly  the  dynamical  timescale at redshift $z$. The
quantity $X$ is defined to be $X\equiv M_{\rm vir}/M_{\rm c}$, where $M_{\rm c}$
is a characteristic mass and $\mathcal{R}$ is a positive number that is  smaller
than $1$. For the star formation rate of satellite galaxies, the related formula
is already provided in Eq.~\ref{eq:sfrsat}.

\begin{figure}
\centering
\includegraphics[width=0.5\textwidth]{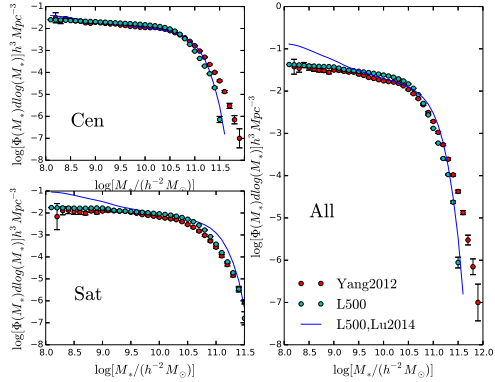}
\caption{The upper left, lower left and right panels show the  galaxy  SMFs  for
  central, satellite  and  all  galaxies, respectively. In  each  panel, the red
  filled  circles  with  error bars are the galaxy stellar mass function of SDSS
  DR7  obtained  by \cite{2012ApJ...752...41Y}. The cyan circles with error bars
  are our fiducial EM results based on L500 simulation. The blue curves  are the
  simular  results  but  based  on  SFH model of \cite{2014MNRAS.439.1294L}. The
  error bars of our EM are calculated using 500 bootstrap re-samplings.}
\label{fig:smf}
\end{figure}

\begin{figure*}
\centering
\includegraphics[width=18.0cm]{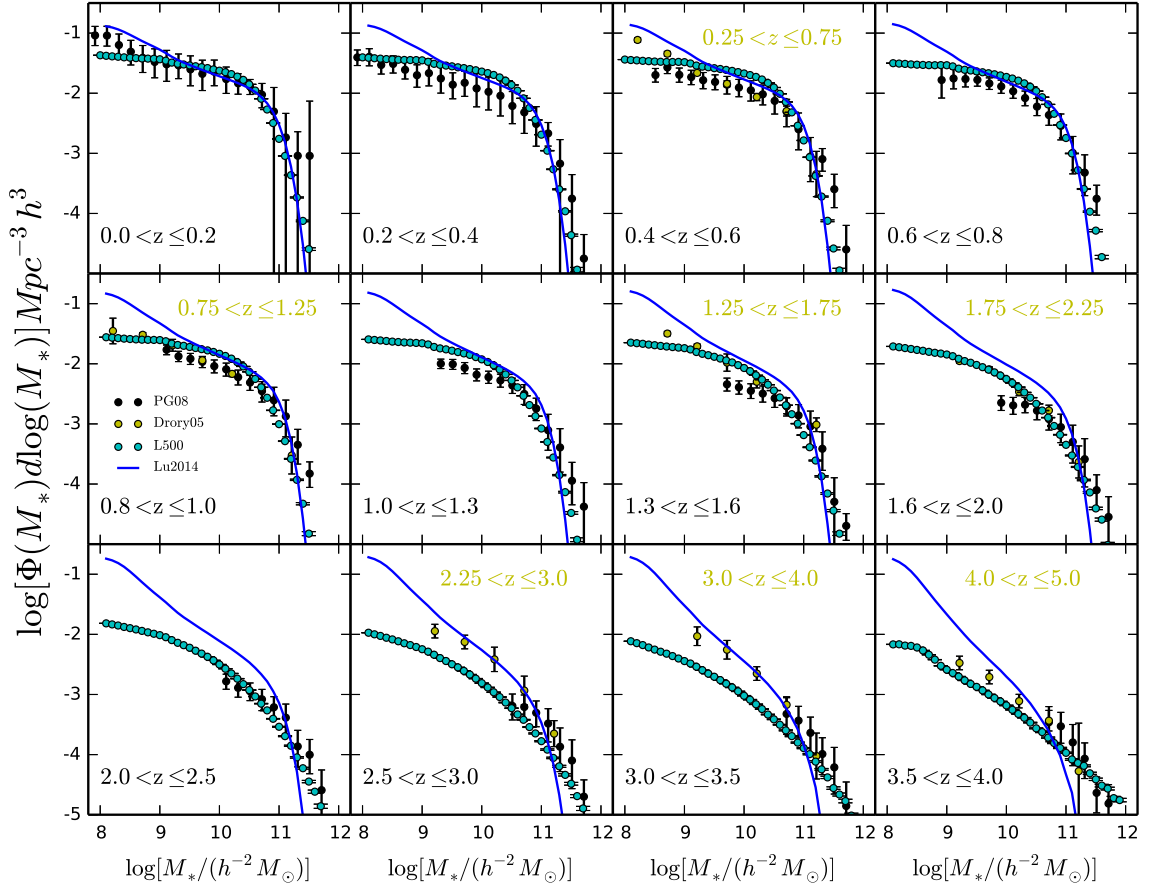}
\caption{Stellar  mass  function  of   galaxies  at  different  redshift bins as
  indicated in each panel. In all panel we compare our EM prediction(cyan filled
  circles) and the Lu14 (blue curve) predictions applied to the L500 simulation,
  to the Spitzer measurements (black circles) published in
  \cite{2008ApJ...675..234P} (PG08). The redshift selection is indicated in black.
  We also add results obtained by \cite{2005ApJ...619L.131D} (Drory05), in similar
  redshift ranges (indicated  in  yellow  for  the relevant panels). are results
  from the Spitzer by \cite{2008ApJ...675..234P} (PG08). The  error  bars of our
  EM are calculated using 500 bootstrap re-samplings.}
\label{fig:highzsmf}
\end{figure*}

\section{The stellar mass properties of galaxies}
\label{sec_result1}

In order to check the performance of our EM for galaxy formation, we  check  the
stellar  mass  function (SMF) and  the  two point correlation function (2PCF) of
galaxies, and compare them to observational measurements. The related observational
measurements are the SMFs at different redshifts (\citealt{2012ApJ...752...41Y};
\citealt{2008ApJ...675..234P}, hereafter PG08; \citealt{2005ApJ...619L.131D},
hereafter Drory05), the CSMFs at low redshift (\citealt{2012ApJ...752...41Y}) and
the 2PCFs for galaxies in different stellar mass bins.

\subsection{SMFs of galaxies at different redshifts}

The  first  set  of observational measurements are the stellar mass functions of
galaxies at redshift $z=0.0$ which are shown in Fig.~\ref{fig:smf} for all (right
panel), central (upper-left  panel)  and  satellite (low-left  panel)  galaxies,
respectively.  The  red circles  with error-bars indicate the observational data
obtained  from  SDSS DR7 by \cite{2012ApJ...752...41Y}.  Cyan circles with error
bars  are  the  results  of our model applied  the halo merger trees of the L500
simulation. Meanwhile, blue curves are obtained using the Lu14 SFH  model on the
same trees.

From the upper-left panel of Fig.~\ref{fig:smf}, it is  clear  that  for central
galaxies the results of our model show an excellent agreement with observational
data within  a  large stellar mass range ($\log \Mstar \sim 8.1-11.0$). However,
in high mass range ($\log\Mstar\ga 11.0$), we somewhat underestimate the stellar
mass function. This discrepancy is probably caused by the fact that in our model,
we used the median SFH to grow galaxies in dark matter halos. However, in reality
scatter of SFHs of high mass central galaxies may be larger  and depend on their
large scale environment. In addition, in our model we did not  take into account
the  major  mergers  of  galaxies, where  only $f_{\rm merger}=0.13$ portion  of
stripped satellite galaxies can be accreted to the central galaxies. For the SFH
models of Lu14,  the results are very similar with our fiducial ones.

For the satellite galaxies, as shown in the lower-left panel of Fig.~\ref{fig:smf},
our  fiducial  EM  reproduces  the  overall  SMFs  quite well. However, a slight
deviation (over prediction) is seen at middle mass range ($\log \Mstar \sim 10.4-10.9$).
In these  satellite  galaxies  either the SFH modelled by Eq.~\ref{eq:sfrsat} is
somewhat  too  strong, or  the stripping and disruption of satellite modelled by
Eq.~\ref{eq:disrupt} is not efficient enough.  As  for  Lu14  model, it does not
match  that  well  with  the SDSS observations, especially in the low mass range
($\log\Mstar \sim 8.0-9.5$). And in high mass range($\log\Mstar\sim 11.0-11.5$),
it over predicts the mass function. Nevertheless, as Lu14 model itself is intended
to  reproduce  the  much  steeper  faint  end  slope of the luminosity function,
especially for satellite galaxies, such differences are expected.

The  right  panel  of  Fig.~\ref{fig:smf}  shows  the  SMF of all galaxies which
include  central galaxies and satellite galaxies. The results of our fiducial EM
in general agree with the observational data, with slight  discrepancies  at the
high mass range ($\log \Mstar \ga 11.0$) mainly  contributed by centrals, and at
middle mass range ($\log\Mstar\sim 10.4-10.9$) mainly contributed by satellites.
The  Lu14  model  show  a larger discrepancy at low mass range($\log \Mstar \sim
8.0-9.5$) which is caused by the satellite components.

Next, we check the stellar mass functions of galaxies at higher redshifts. Shown
in Fig.~\ref{fig:highzsmf} are  SMFs  of  galaxies at different redshift bins as
indicated in each panel. In these higher  redshift  bins, in order to mimics the
typical error in the stellar mass estimation in observations, we add logarithmic
scatters to the stellar masses of galaxies as $\sigma_c(z) = {\rm max} [0.173, 0.2
z]$ (see \citealt{2012ApJ...752...41Y} for more detail). The yellow filled circles
with  error-bars  are  results  obtained by \cite{2005ApJ...619L.131D}, in which
they have combined  the data from FORS Deep and from the GOODS/CDFS Fields.  The
cyan circles with error bars are our EM results based on L500  simulation, while
blue curves are the results of Lu14 model based on L500 simulation.

As shown in Fig.~\ref{fig:highzsmf}, in both low and high redshift bins  $z<1.0$
and $z>2.0$, the  SMFs  from  our  model agree quite well with the observational
results.  However in the redshift range $1.0<z<2.0$, our model over predicts the
SMFs. As seen in the lower-left  panel of Fig.~\ref{fig:smf}, this discrepancies
might be due to some over prediction of satellite galaxy counts.  In comparison,
we also show results based  on Lu14 model, which present even higher SMFs within
the redshift range $1.0<z<2.0$.

\begin{figure*}
\centering
\includegraphics[width=18.0cm]{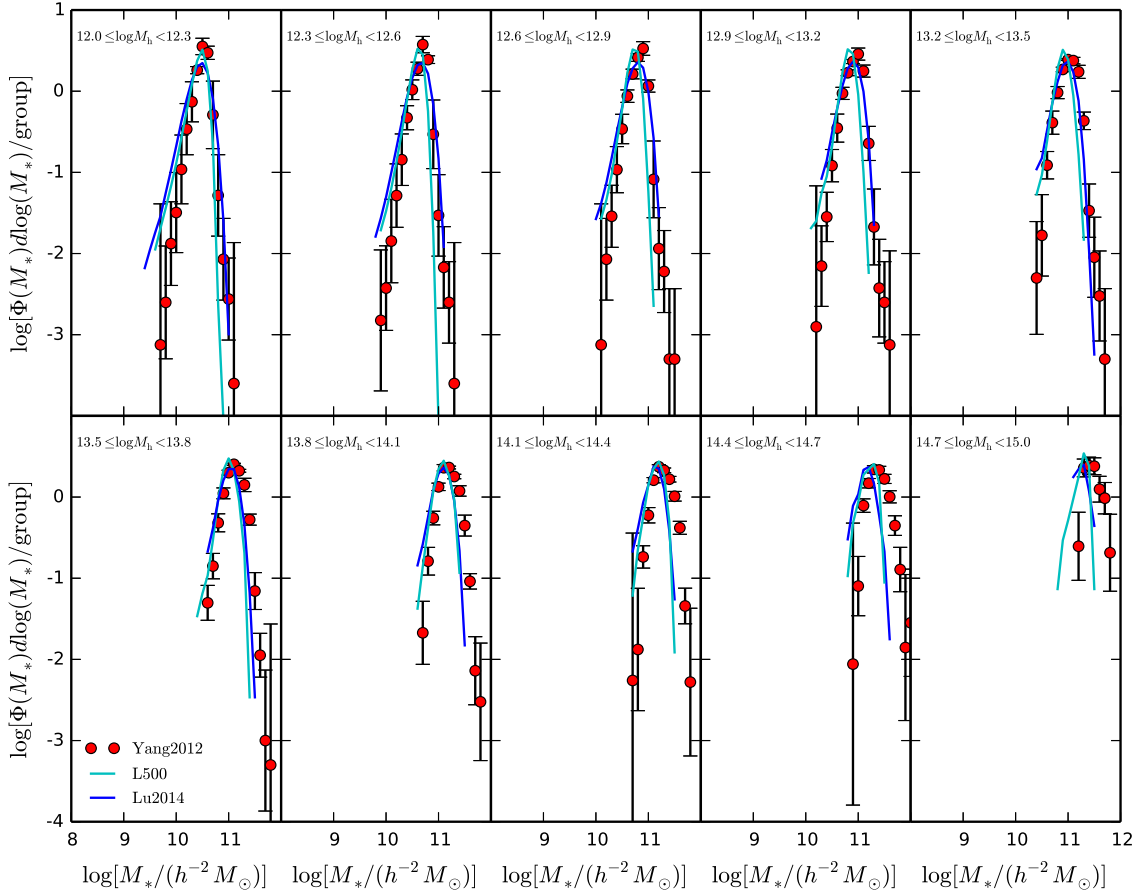}
\caption{Conditional stellar mass functions (CSMFs) of central galaxies. Different
  panel  corresponds  to  different  halo  mass bin as indicated. The cyan solid
  curves are the CSMFs of our EM, while blue curves are obtained  using Lu14 SFH
  model. Red filled circles with error bars  are  the CSMFs of SDSS DR7 obtained
  by \cite{2012ApJ...752...41Y}.  }
\label{fig:csmf_cen}
\end{figure*}

\begin{figure*}
\centering
\includegraphics[width=18.0cm]{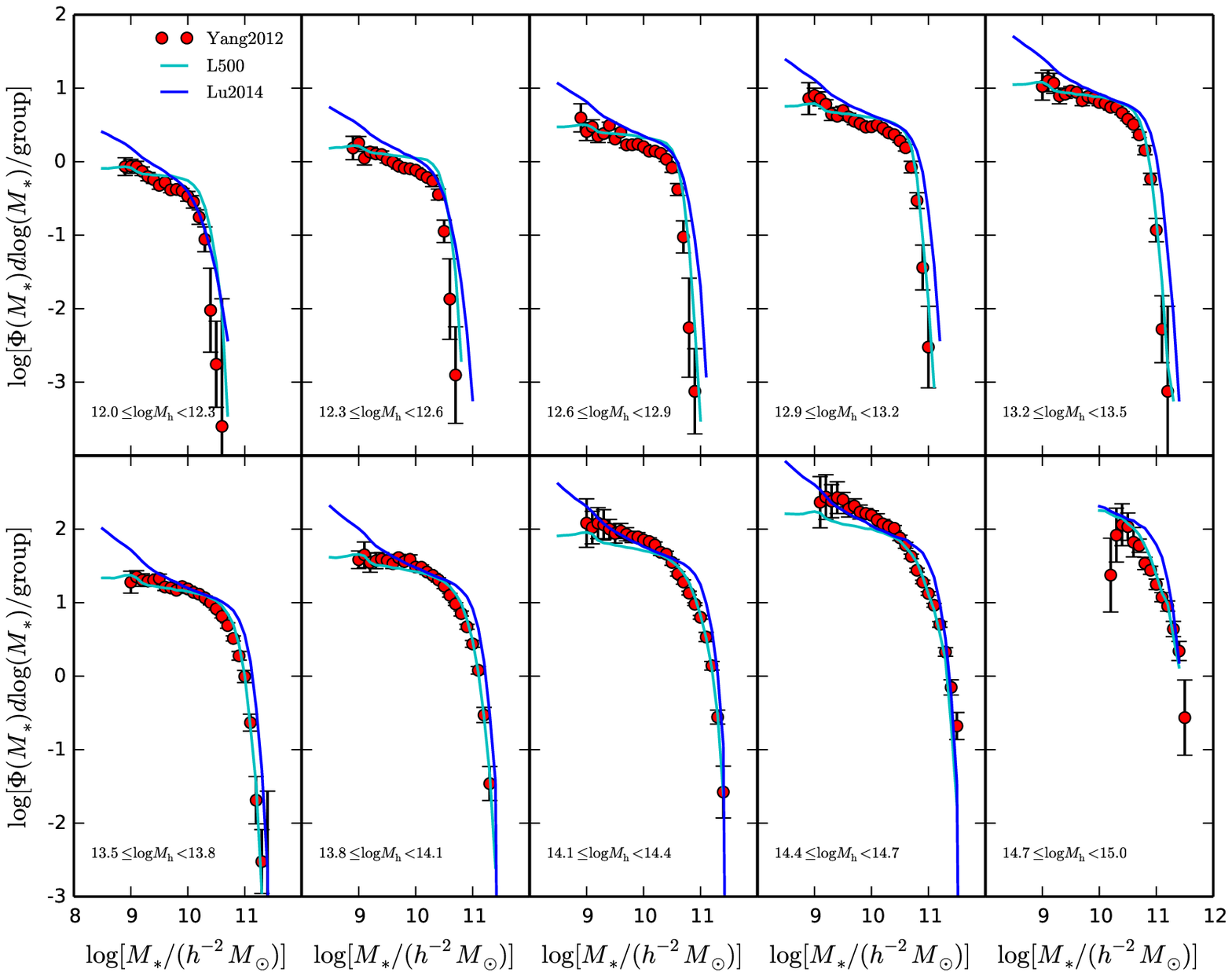}
\caption{Similar to Fig.~\ref{fig:csmf_cen} but for satellite galaxies.}
\label{fig:csmf_sat}
\end{figure*}

\subsection{CSMFs of galaxies at $z=0$}

The conditional stellar mass function (CSMF) $\phi(\Mstar|\Mh)$, which describes
the  average  number  of  galaxies as a function of galaxy stellar mass $\Mstar$
that  can be formed within halos of mass $\Mh$, is an important measure that can
be used to constrain galaxy formation models. As carried out in
\cite{2010ApJ...712..734L} using  the  CSMFs  of  satellite  galaxies, classical
semi-analytical  models  at  that  time  typically  over predicted the satellite
components  by  a  factor  of  two which indicates that either less (or smaller)
satellites can be formed, or more  satellite galaxies need to be disrupted. Here
we compare our model predictions with observational data in Fig.~\ref{fig:csmf_cen}
and Fig.~\ref{fig:csmf_sat} for central and satellite galaxies separately.

Based on the SDSS DR7 galaxy  group catalog, \cite{2012ApJ...752...41Y} obtained
the CSMFs of central galaxy and satellite galaxies, which are shown  as  the red
filled circles with error-bars in Fig.~\ref{fig:csmf_cen} and Fig.~\ref{fig:csmf_sat},
respectively.  The CSMFs from our model are  shown  as  cyan solid curves.  Blue
curves are the CSMFs obtained from galaxy catalogs constructed using Lu14 model.
As  shown  in Fig.~\ref{fig:csmf_cen}, the central galaxy CSMFs of our model and
Lu14  model  are very similar.  Both of them agree well with the observations in
halo  mass  range $12.0 \le \log \Mh < 13.8$ but are slightly under estimated in
halo mass range $13.8 \le \log \Mh < 15.0$. As  shown in Fig.~\ref{fig:csmf_sat}
for satellite galaxies, the  CSMFs  of our model agree well with the observation
in general. There are little deviations in halo mass ranges $12.0\le\log\Mh<12.3$,
$12.3 \le \log \Mh < 12.6$ and $12.6 \le \log \Mh < 12.9$. In these  ranges, our
model  overestimates  the  CSMFs  at $9.5 \le \log \Mstar < 10.5$. Thus the over
predicted  satellite  galaxies  shown  in Fig.~\ref{fig:smf} are mainly in these
Milky  Way  sized  and  group sized halos.  While in Lu14 model, as seen for the
satellite galaxy stellar mass function shown in Fig.~\ref{fig:smf}, the CSMFs in
halos of different masses all show an upturn at low mass end.

\begin{figure}
\centering
\includegraphics[width=9.0cm]{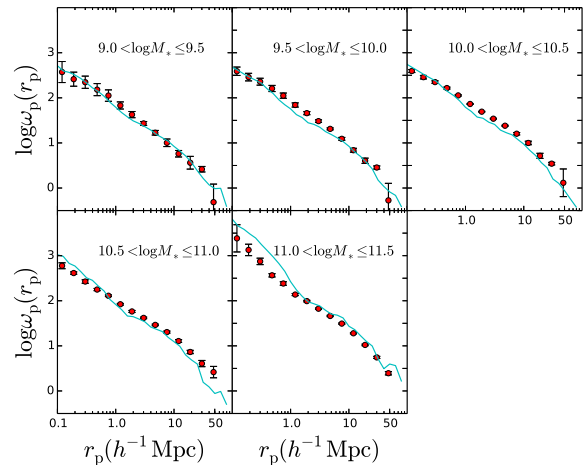}
\caption{Projected 2PCFs of galaxies in different stellar mass bins as indicated
  in each panel. Red filled circles  with  error  bars are the 2PCFs of SDSS DR7
  obtained by \cite{2012ApJ...752...41Y} and cyan curves are our EM results.}
\label{fig:2pcf}
\end{figure}

\subsection{2PCFs of galaxies}

The two point correlation function which measures the excess of galaxy  pairs as
a  function  of  distance  is  a widely used quantity to describe the clustering
properties of galaxies. In terms of galaxy formation, it can be used to constrain
the HOD of galaxies (\citealt{1998ApJ...494....1J}) and  to constrain the CLF of
galaxies (\citealt{2003MNRAS.339.1057Y}).  Here we compare the model predictions
of 2PCFs in our galaxy catalogs to observations.

Fig.~\ref{fig:2pcf} shows the projected 2PCFs of  galaxies  in different stellar
mass  bins. Our  model  predictions  are  shown  as  the  solid  curves  and the
observational data obtained by \cite{2012ApJ...752...41Y} from SDSS DR7 are shown
as the filled circles with error bars. Our overall model predictions are quite a
good match to the observation in the stellar mass range $9.0<\log\Mstar < 11.0$.
However, in the most massive stellar mass bin ($11.0 < \log \Mstar < 11.5$), our
model results is higher than the observations for $r_{\rm p}\la 1h^{-1}\rm Mpc$.
The  too  strong  clustering  at $r_{\rm p}<1\mpch$ for  these high mass objects
is  mainly  caused  by  the  fact that due to the insufficient prediction of the
central  galaxies, the  satellite  fraction  in  this mass bin is over predicted
(see Fig.~\ref{fig:smf}).

\begin{figure}
\centering
\includegraphics[width=9.0cm]{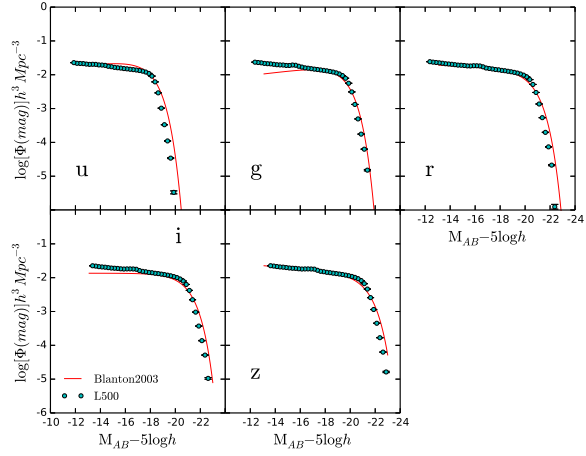}
\caption{Luminosity  functions  of  galaxies in $u, g, r, i, z$ band at $z=0.1$.
  The solid curve in each panel is the corresponding best fit  Schechter form LF
  obtained by \cite{2003ApJ...592..819B} from SDSS DR1.
}
\label{fig:all_lf}
\end{figure}

\begin{figure}
\centering
\includegraphics[width=9.0cm]{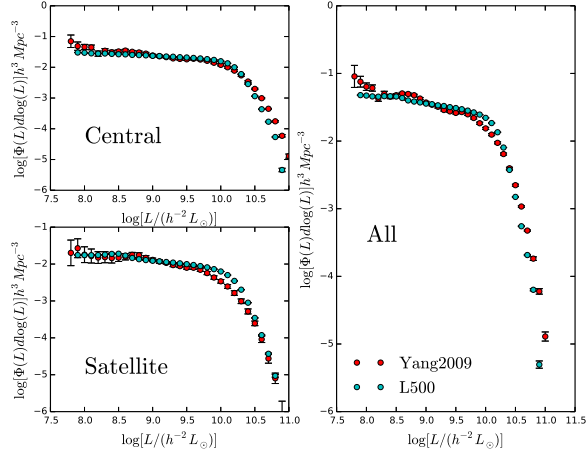}
\caption{Luminosity functions of central, satellite and all galaxies in $r$ band
  in the local universe. Here results  are  shown for observational measurements
  (red dots) and our fiducial model predictions (cyan dots), respectively.  }
\label{fig:lf_yang}
\end{figure}

\begin{figure*}
\centering
\includegraphics[width=18.0cm]{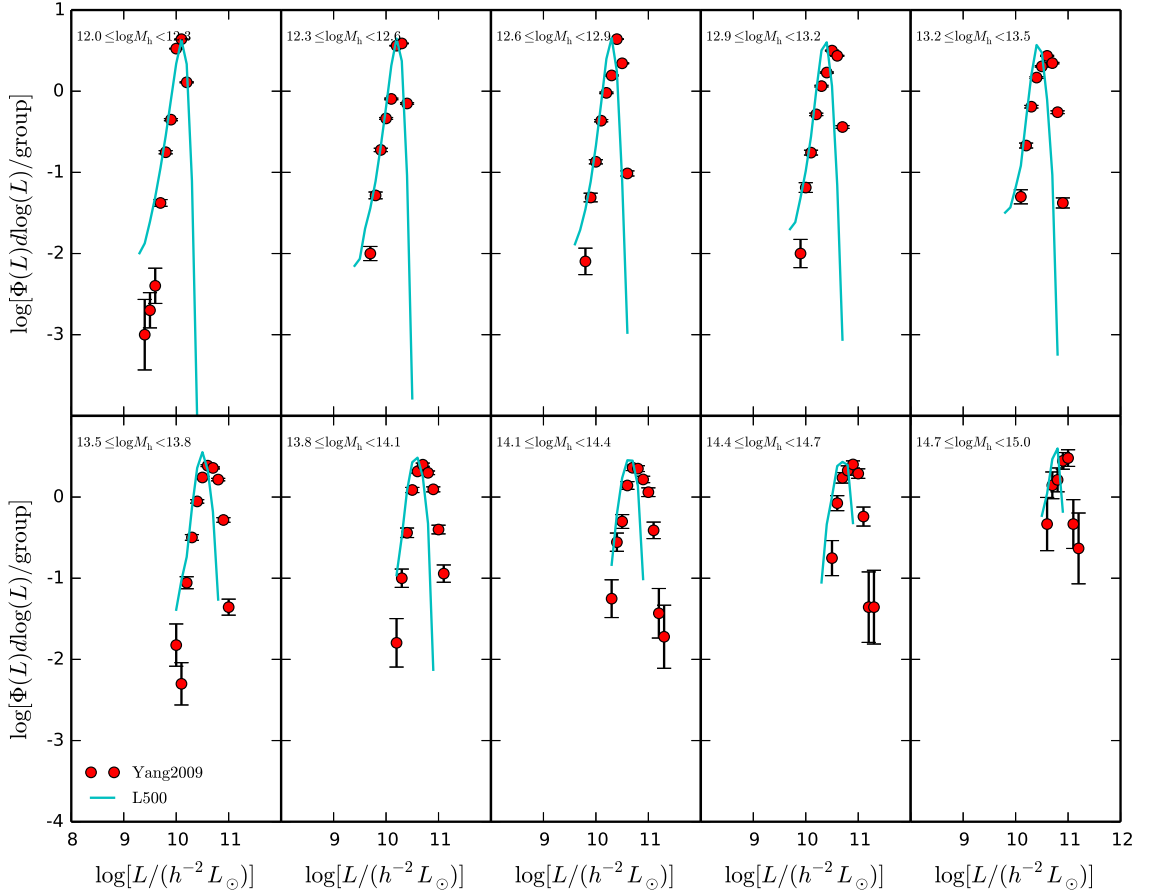}
\caption{Conditional luminosity functions (CLFs) of central galaxies in halos of
  different  mass  bins. Here  results  are shown for observational measurements
  (red dots) and our fiducial model predictions (cyan curves), respectively.  }
\label{fig:clf_cen}
\end{figure*}

\begin{figure*}
\centering
\includegraphics[width=18.0cm]{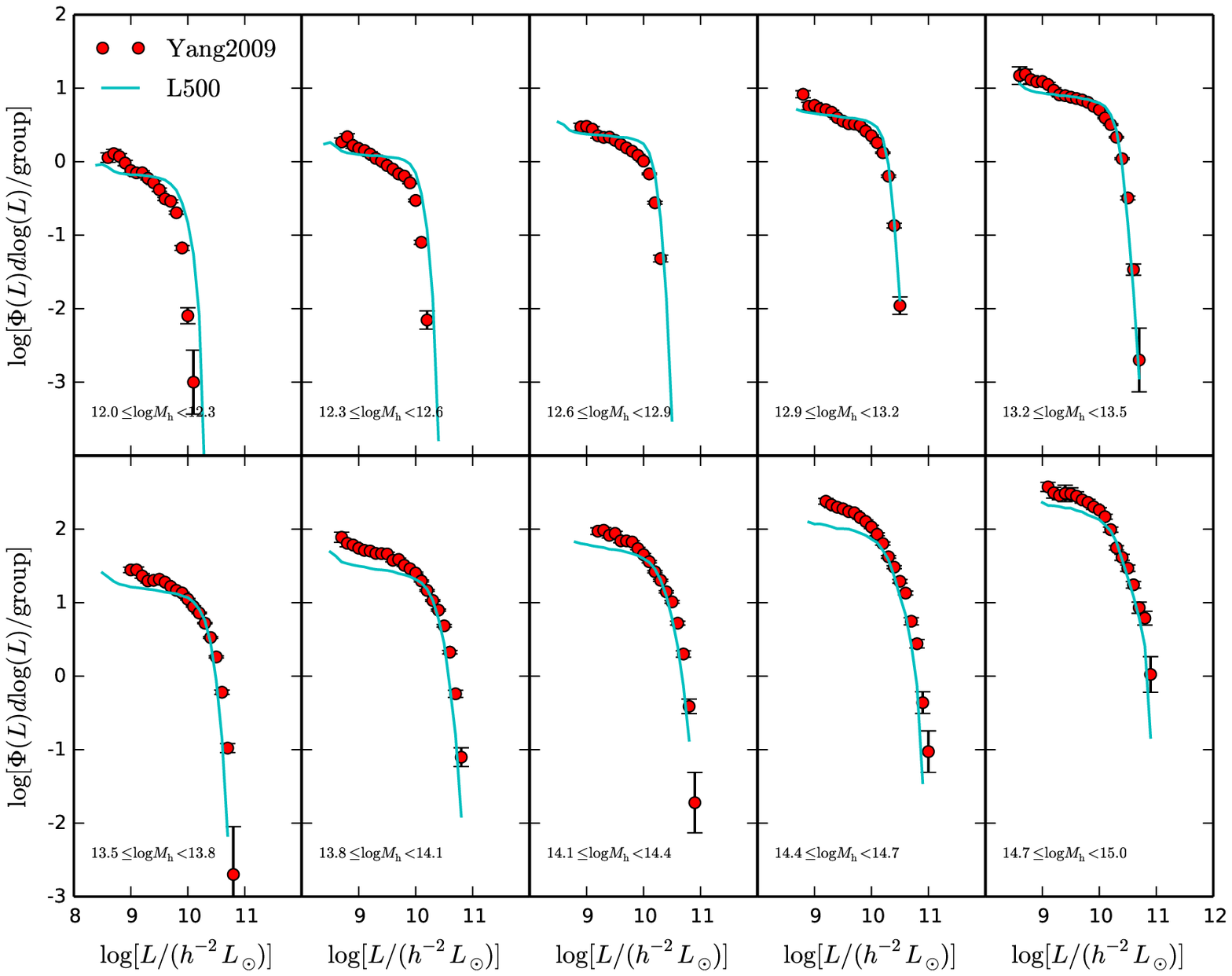}
\caption{Similar as Fig.~\ref{fig:clf_cen} but for the satellite galaxies.}
\label{fig:clf_sat}
\end{figure*}

\section{The luminosity and gas properties of galaxies}
\label{sec_result2}

Apart from the stellar masses of galaxies, we now turn to the luminosity and gas
components of galaxies.

\subsection{Luminosities of galaxies in different bands}

As  detailled  in  section~\ref{sec_sfr}, from the halo merger histories derived
from  the  L500  simulation, we model galaxies from a estimation of their sellar
mass  and  SFR  as  a  function of time. We use those information to predict the
photometric  properties  of  our  model  galaxies  using  the stellar population
synthesis model of \cite{2003MNRAS.344.1000B} using a Salpeter IMF
(\citealt{1955ApJ...121..161S}). Since    our   model   does not include the gas
component in galaxies, we cannot  directly  trace the  chemical evolution of the
stellar population. To circumvent this problem,  we follow the metallicity - stellar
mass  relation   derived  in  Lu14 from observation of galaxies at all redshifts
{\it specified redshift range of ranges}. We adopt  the  mean  relation based on
the data of \cite{2005MNRAS.362...41G}, which can roughly be described as
\begin{equation}
 \log_{10} Z = \log_{10} Z_{\odot} + \frac{1}{\pi}
\tan \left[\frac{\log_{10}(M_{\star}/10^{10}M_{\odot})}{0.4}\right] - 0.3 \,.
\end{equation}
This  observational  relation  extends down to a stellar mass of $10^9M_{\odot}$
and has a scatter of $0.2\,{\rm dex}$ at the massive end and of $0.5\,{\rm dex}$
at the low mass end.

Using the stellar population synthesis model, we can obtain galaxy  luminosities
in different bands. We show in Fig.~\ref{fig:all_lf}, the  luminosity  functions
of  all  galaxies in the five different SDSS bands ($u, g, r, i, z$) at $z=0.1$.
For  comparison, we  also  show  in  each panel the corresponding best Schechter
functional  LFs  fit  obtained  by \cite{2003ApJ...592..819B} from SDSS DR1. The
observational measurements and corresponding model fitting are  roughly  limited
to absolute magnitude limit ($-16, -16.5, -17, -17.5, -18$) in ($u, g, r, i, z$)
bands, respectively. Within these magnitude limits, our  model predictions agree
with  the  observational  data fairly well with very slight under predictions at
the bright ends. Only in $u$-band do we see a pro-eminent deficit of galaxies at
$\umag\sim -16.0$. These behaviors indicate that the  stellar  compositions as a
function of time as derived with our model are on average accurate.

In  addition  to  the LFs  of the full galaxy population, we can distinguish the
contribution from the centrals and  the satellites. Fig.~\ref{fig:lf_yang} shows
the $r$ band luminosity functions of all (right panel), central (upper-left panel)
and  satellite (low-left panel) galaxies. Our  fiducial  model  predictions  are
shown as the cyan dots with error bars obtained from 500 bootstrap re-samplings.
Red  points  with error bars are obtained by \cite{2009ApJ...695..900Y} but were
updated to SDSS DR7. Similar with Fig.~\ref{fig:smf}, our  model  underestimates
the central galaxy luminosity function at high luminosity end ($10.5 \la \log L
\la 11.0$) and  overestimates  the  satellite  galaxy luminosity function in the
luminosity range ($10.0\la \log L \la 10.5$).

Similarly to the CSMFs, the conditional  luminosity functions(CLFs) describe, as
a  function  of  luminosity $L$, the  average  number of galaxies that reside in
dark  matter  halo  of  a given mass $M_{h}$. In Fig.~\ref{fig:clf_cen} the CLFs
obtained  from  our mock catalogs are compared to the observational measurements
obtained by \cite{2009ApJ...695..900Y}(also updated to SDSS DR7).  As one  could
expect, the  performances  of  CLFs  of central galaxies is quite similar to the
situation found for the CSMFs in Fig.~\ref{fig:csmf_cen}. The central galaxy CLFs
of our model agree well with the observational results in the $12.0 \le \log \Mh
<13.5$ halo mass range still there is some discrepancy for $13.5\le\log\Mh<15.0$.

As for the satellite galaxies shown in Fig.~\ref{fig:clf_sat}, the  situation is
somewhat  different  with  respect  to  the  CSMFs. Our  model matches well with
observations in $12.9 \le \log \Mh < 13.8$, while  underestimate  the  number of
satellite galaxies at the low luminosity end in high mass halos $13.8\le\log \Mh
< 15.0$. These  discrepancies are highly interesting as they differ from the one
we found for the CSMFs (Fig \ref{fig:csmf_sat}), as it indicates that the colors
of these galaxies are not entirely properly modelled.

\begin{figure}
\centering
\includegraphics[width=9.0cm]{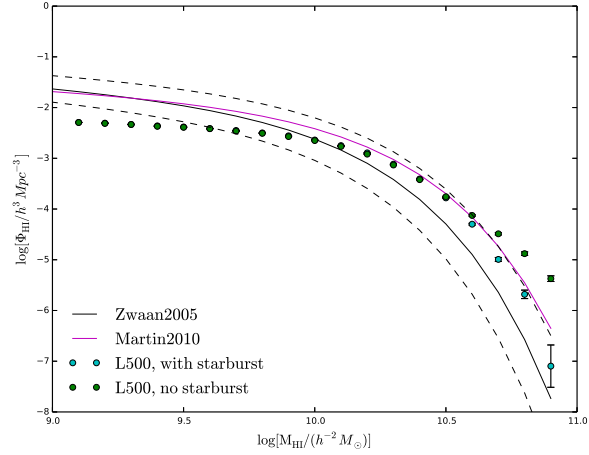}
\caption{The HI mass functions: green dots are our fiducial  model  predictions,
  while cyan dots are the model predictions that taken into account the starburst.
  The  black  solid  line  shows  the best fit observational results obtained by
  \cite{2005MNRAS.359L..30Z} with dashed lines indicate its $\pm1\sigma$ scatter.
  Magenta curve is the observational fitting formula obtained by
  \cite{2010ApJ...723.1359M}.}
\label{fig:himf}
\end{figure}

\begin{figure}
\centering
\includegraphics[width=9.0cm]{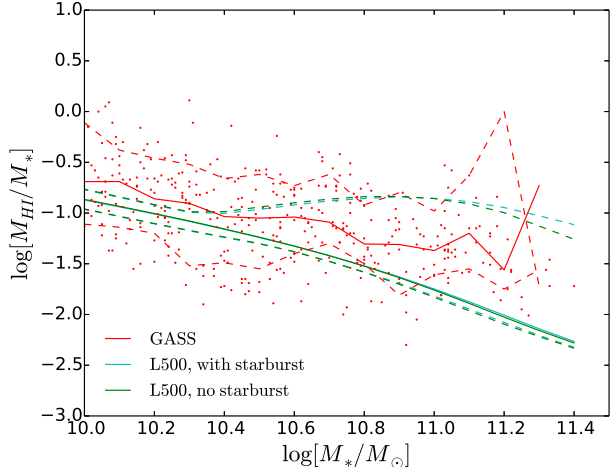}
\caption{The  HI-to-stellar  mass  ratios  as a function of galaxy stellar mass.
  Here red points are data from GASS (\citealt{2013MNRAS.436...34C}) survey. Red
  curves are the median and 68\% confidence range of the ratio in the GASS sample.
  Green  and  cyan  curves represent the median and 68\% confidence range of our
  fiducial and star-burst model predictions, respectively.  }
\label{fig:hi_frac}
\end{figure}

\subsection{{\rm HI} masses of galaxies}

Although  our  EM  is  limited  to model the star components of galaxies, we can
estimate  the  gas components within the galaxies. Here we focus on the cold gas
that are associated with the star formation(\citealt{1959ApJ...129..243S}).  The
star formation law most widely implemented in SAM was proposed by
\cite{1998ApJ...498..541K} as follows:
\begin{eqnarray}\label{eq:sfr}
\Sigma_\mathrm{SFR} & = & (2.5 \pm 0.7) \times 10^{-4} \nonumber \\
			 &   &	(\frac{\Sigma_\mathrm{gas}}{1 \Msun \mathrm{pc}^{-2}})^
			{1.4 \pm 0.15}\Msun \mathrm{yr}^{-1} \mathrm{kpc}^{-2}	\, ,
\end{eqnarray}
where $\Sigma_\mathrm{SFR}$ and $\Sigma_\mathrm{gas}$ are the surface  densities
star formation and gas, respectively.

In  this  paper, we  use  the  model  proposed  in \cite{2010MNRAS.409..515F} to
estimate the cold gas within our galaxies. This method consists in following the
build-up of stars and gas within a fixed set of 30 radial ``rings''. The  radius
of each ring is given by the geometric series
\begin{eqnarray}\label{eq:ri}
r_{i} & = & 0.5 \times 1.2^{i}[h^{-1} \rm{kpc}](i=1,2...30)	.
\end{eqnarray}
According to \cite{1998MNRAS.295..319M}, the cold gas is distributed exponentially
with surface density profile
\begin{eqnarray}\label{eq:sigma_gas}
\Sigma_\mathrm{gas} (r) & = & \Sigma_\mathrm{gas}^\mathrm{0} \exp
(-r/r_\mathrm{d})\,,
\end{eqnarray}
where $r_\mathrm{d}$ is the scale length of the galaxy, and
$\Sigma_\mathrm{gas}^\mathrm{0}$ is given by $\Sigma_\mathrm{gas}^\mathrm{0}
=  m_\mathrm{gas} / (2 \pi r_\mathrm{d}^\mathrm{2})$.

With the above ingredients, we are able to predict the total amount of  cold gas
associated to each galaxy. However, observationally, we only  have  a relatively
good  estimate  of  HI  mass  in the local universe. Here we calculate HI masses
associated to galaxies by assuming a constant ${\rm H}_{2}/{\rm HI}$ ratio of 0.4
and a hydrogen mass fraction $X=0.74$ (\citealt{2011MNRAS.416.1566L};
\citealt{2004MNRAS.351L..44B}; \citealt{2010MNRAS.406...43P}). Fig.~\ref{fig:himf}
shows the HI mass function of  galaxies  in the local universe obtained from our
mock galaxy catalog (green dots). For comparison, we also show in Fig.~\ref{fig:himf},
using black curve, the fitting formula of HI mass function obtained by
\cite{2005MNRAS.359L..30Z} from HIPASS:
\begin{eqnarray}\label{eq:hi}
\Theta(M_\mathrm{HI})dM_\mathrm{HI} =
	\left(\frac{M_\mathrm{HI}}{M^{\ast}_\mathrm{HI}}\right)^{\alpha}
	\exp \left(-\frac{M_\mathrm{HI}}{M^{\ast}_\mathrm{HI}}\right)
	d \left(\frac{M_\mathrm{HI}}{M^{\ast}_\mathrm{HI}}\right)  \,   ,
\end{eqnarray}
where $\alpha=-1.37 \pm 0.03$ and $\log (M^{\ast}_\mathrm{HI}) /\Msun = 9.80 \pm
0.03 h^{-2}_{75}$. Black  dashed  lines  indicate  the $\pm1\sigma$  scatter. An
additional observational HI mass function is obtained by \cite{2010ApJ...723.1359M}
using $1/V_{max}$ method (magenta curve).

Our model only shows a fair agreement  to  these observational data, even though
it under predicts the HI mass function at $\log \Mhi \la 9.6$ and  over predicts
the HI mass function at $\log \Mhi \ga 10.5$. These  discrepancies  are possibly
caused by different factors. The  first  one am be, of course, the uncertainties
in  the  SFR-cold  gas  mass ratios. In addition to this, as the SFR in low mass
halos have much larger scatters than the ones we adopt here (see Fig. 1 in
\cite{2013ApJ...770..115Y}), adopting  a larger scatter may help to solve the HI
mass  function  deficiency  at  low mass end.  On the massive end of the HI mass
function, the difference may be connected to starburst galaxies (with high SFR).
However, in reality, starburst is not necessary associated with the largest cold
gas component.  As \cite{2014ApJ...789L..16L} have  checked  the morphologies of
star-burst  galaxies (with  SFRs 5 time  higher  than  the  median for the given
stellar  mass), and  found  that  more than half of them are associated with gas
rich major mergers.  To partly take this into account, we adopt the  collisional
star-burst model proposed by \cite{2001MNRAS.320..504S} used in many SAMs
(\citealt{2008MNRAS.391..481S}; \citealt{2011MNRAS.413..101G}). During the
star-burst process, the increased stellar mass of the central galaxy is

\begin{eqnarray}\label{eq:burst}
\delta m_\mathrm{starburst} & = & (m_\mathrm{gas,\;sat}+m_\mathrm{gas,\;cen})\nonumber \\
    &	& e_\mathrm{burst}  \left (\frac{m_\mathrm{sat}}{m_\mathrm{cen}}\right)^\mathrm{\gamma_{burst}} \, ,
\end{eqnarray}

where $m_\mathrm{gas,\;cen}$ ($m_\mathrm{gas,\;sat}$) is  the  cold  gas mass of
central (satellite)  galaxy, $m_\mathrm{cen}$ ($m_\mathrm{sat}$) is  the  sum of
stellar mass and cold gas mass of central (satellite) galaxy, $e_\mathrm{burst}=
0.55$, and $\gamma_\mathrm{burst}=0.69$. The values of $e_\mathrm{burst}$ and
$\gamma_\mathrm{burst}$ are determined from isolated galaxy merger
simulations performed by \cite{2008MNRAS.384..386C}. Within our merger trees, we
identify these star-burst  galaxies  and  swap their SFRs to the highest ones in
similar  mass  halos. The  cold  gas  for  these galaxies are updated using this
Eq.~\ref{eq:burst}. We  show  in  Fig.~\ref{fig:himf} using  cyan  dots how this
starburst  implementation  successfully  corrects the over-estimation of HI mass
function at massive end.

Apart from the HI mass functions, we also compare the HI-to-stellar  mass ratios
of  galaxies. Fig.~\ref{fig:hi_frac}  illustrates  the  HI-to-stellar mass ratio
$\log[M_{\rm HI}/M_{\ast}]$ as a function of galaxy stellar mass. Red points are
from GASS survey (\citealt{2013MNRAS.436...34C})  while the red curve represents
the  median  value  and  red  dashed curves indicate the $16^{th}$ and $84^{th}$
percentile  ranges  of  $\log[M_{\rm HI}/M_{\ast}]$. The  green solid and dashed
curves represent the median and $16^{th}$ and $84^{th}$ percentile ranges of our
fiducial model prediction from the L500 simulation.  While the cyan  curves  are
obtained  from  the  star-burst variation of the model. We can see that both our
models reproduce the average trends of HI-to-stellar  mass  ratios as a function
of stellar mass quite well. But  the  scatter of the model prediction is smaller
than  the  observation  at  low  masses. We  think  that  this may caused by the
relation between star formation rate and cold gas used in our model.

%%%%%%%%%%%%%Discussion%%%%%%%%%%%%%%%%%%%%%%%%%%%%%%%%

\section{Summary}
\label{sec_dis}

Based  on  the star formation histories of galaxies in halos of different masses
derived by \cite{2013ApJ...770..115Y}, we an empirical model to study the galaxy
formation and evolution. Compared to traditional SAMs, this  model  has few free
parameters, each of which can be associated with the observational data. Applying
this model to merger trees derived from $N$-body simulations, we predict several
galaxy properties that agree well with the observational data.  Our main results
can be summarized as follows.

\begin{enumerate}

\item At redshift $z=0$, the SMFs of all galaxies agree well with the observation
  within $8.0<\log\Mstar<11.3$ but our estimate is slightly low in high  stellar
  mass end ($11.3 < \log \Mstar < 12.0$).

\item Our  SMFs  show  generally a fair agreement with the observational data at
  higher redshifts up to 4.  While in redshift $1.0<z<2.0$, the SMFs  at the low
  mass end are somewhat over-estimated.

\item At  redshift  $z=0$, the  CSMFs  of  central  galaxies agree well with the
  observations in the $12.0\le\log\Mh<13.8$ halo mass range and somewhat shifted
  to lower masses in halo mass range $13.8 \le \log\Mh < 15.0$. In the meantime,
  the CSMFs of satellite galaxies agree quite well with the observations.

\item The  projected  2PCFs  in  different stellar mass bins calculated from our
  fiducial  galaxy  catalog  can  match  well the observations. Only in the most
  massive stellar mass bin the correlation is over predicted at small scales.

\item We can derive  from our model LFs  in the $^{0.1}u$, $^{0.1}g$, $^{0.1}r$,
  $^{0.1}i$ and $^{0.1}z$ bands. They  prove  to  be roughly consistent with the
  SDSS observational results obtained by \cite{2003ApJ...592..819B}.

\item The  central  galaxy  CLFs  of our model agree well with the observational
  results in halo mass range $12.0\le\log\Mh < 13.5$, quite similar to the SMFs.
  However, the satellite galaxy CLFs are somewhat underestimated at faint end in
  halos with mass $12.9 \le \log \Mh < 13.8$.

\item Our  prediction  of  HI mass function agree with the observational data at
  roughly $\pm1\sigma$ level at $\log \Mhi \ga 9.6$, and somewhat underestimated
  at lower mass ends.

\end{enumerate}

Our  model  predicts  roughly  consistent, although  not  perfect, stellar mass,
luminosity and HI mass components of galaxies. Such a method is a potential tool
to   study  the  galaxy  formation  and  evolution  as an alternative to SAMs or
abundance matching methods. The galaxy and gas  catalogs here constructed can be
used to construct redshift surveys for future deep surveys.

\normalem
\begin{acknowledgements}
This  work  is  supported  by  973  Program (No. 2015CB857002), national science
foundation of China (grants Nos. 11203054, 11128306, 11121062, 11233005, 11073017,
11421303), NCET-11-0879, the Strategic Priority Research Program ``The Emergence
of Cosmological Structures" of  the  Chinese   Academy   of  Sciences, Grant No.
XDB09000000 and the Shanghai Committee  of  Science and Technology, China (grant
No. 12ZR1452800). SJL thanks Ming Li for his help in dealing with the simulation
data, Ting  Xiao for her useful discussion concerning HI gas and Jun Yin for her
help in stellar population synthesis modeling.

A computing facility award on the PI cluster at Shanghai Jiao Tong University is
acknowledged. This  work  is  also  supported  by the High Performance Computing
Resource in the Core Facility for Advanced Research Computing at Shanghai 
Astronomical Observatory.

\end{acknowledgements}

\bibliographystyle{raa}
\bibliography{bibtex}

\end{document}